\documentclass[11pt,a4paper]{article}

\date{}

\usepackage{amsfonts,amscd,amssymb,amsmath,amsthm,latexsym}
\usepackage{verbatim}
\usepackage{graphicx, subfigure,color}

\date{}

\begin{document}

\title{Cosmological  Solutions of a  Nonlocal Square Root Gravity }

\author{I. Dimitrijevic$^a$, B. Dragovich$^{b,c}$, A. S. Koshelev$^d$, \\ Z. Rakic$^a$ and J. Stankovic$^e$ \\
$^a$Faculty of Mathematics, University of Belgrade,
Studentski trg 16, \\ Belgrade, Serbia    \\
$^b$Institute of Physics, University of Belgrade,   Belgrade, Serbia \\
$^c$Mathematical Institute, Serbian Academy of Sciences and Arts,\\ Belgrade, Serbia \\
$^d$Departamento  de F\'isica and Centro  de  Matem\'atica  e
Aplica\c c\~oes, \\  Universidade  da  Beira  Interior,  6200  Covilh\~a,
Portugal  \\
$^e$Teacher Education Faculty, University of Belgrade,  Kraljice Natalije 43,\\ Belgrade, Serbia}


\maketitle



\begin{abstract}
In this paper we consider modification of general relativity extending $R - 2 \Lambda$ by nonlocal term of the form  $\sqrt{R-2\Lambda}\, \mathcal{F}(\Box)\,  \sqrt{R-2\Lambda} ,$ where $\mathcal{F}(\Box)$ is an analytic function of the d'Alembert operator $\Box$. We have found some exact cosmological solutions of the corresponding equations of motion without matter and with $\Lambda \neq 0$. One of these solutions is $ a (t) = A \, t^{\frac{2}{3}} \, e^{\frac{\Lambda}{14} t^2} ,$ which imitates  properties similar to an interplay of the dark matter and the dark energy. For this solution we calculated some cosmological parameters which are in a good agreement with observations. This nonlocal gravity model has not the Minkowski space solution.  We also found several  conditions which function $\mathcal{F}(\Box)$ has to satisfy.
\end{abstract}

\section{Introduction}

Despite of numerous significant phenomenological confirmations and many nice theoretical properties, General relativity (GR) \cite{wald} is not final theory of gravity.
Problems mainly come from quantum gravity, cosmology and astrophysics. For example, if GR (as Einstein theory of gravity) is applicable to the  universe as a whole and the universe is homogeneous and isotropic, then it follows that the universe contains about 68 \% of  dark energy (DE), 27 \% of  dark matter (DM) and only about 5 \% of  visible matter.   However, validity of GR at the cosmological scale is not verified, as well as DE and DM are not yet observed in laboratory experiments. Also, from GR follows cosmological singularity. These and other problems give rise to extensions of GR.    Note that there is no firm theoretical principle which could tell us in which direction to look for a solution. Therefore, there are many attempts to modify GR, e.g.  see review articles \cite{faraoni,clifton,nojiri,nojiri1,novello}. One of actual approaches to modification of GR in domain of cosmology is nonlocal modified gravity, see e.g. \cite{deser,woodard,dragovich0,maggiore,biswas1,biswas2,biswas3,biswas4,biswas5,koshelev,koshelev1,koshelev2,koshelev3,koshelev4,eliz,koivisto}.

All nonlocal gravity models contain the gravitational d'Alembert operator $\Box$, which mainly appears in two ways: 1) in the form of an analytic function
$F (\Box) = \sum_{n=0}^{+\infty} f_n \Box^n$ and 2) in the form $\Box^{-1}$. Models with inverse d'Alembert operator are considered as possible explanation of the late cosmic acceleration without use of a dark matter.  Usually the form of such models is
\begin{equation} \label{1.1}
S = \frac{1}{16\pi G} \int \sqrt{-g} \left(  R + L_{nl}  \right) \, d^4x ,
\end{equation}
where, for examples:  $L_{nl} = R\, f(\Box^{-1}R)$ (see  \cite{nojiri,woodard,koivisto} and references therein), and  $ L_{nl} = -\frac{1}{6} m^2 R \Box^{-2} R $ (see \cite{maggiore} and references therein), and $R$ is the scalar curvature.

Usage of $F (\Box) = \sum_{n=0}^{+\infty} f_n \Box^n$ is influenced by existence of analytic expressions with  $\Box$ in string field theory and $p$-adic string theory (see \cite{dragovich1} and references therein). Some modified gravity models with analytic nonlocality, which have been so far considered, are  concrete examples of the action (without matter)
\begin{equation} \label{1.2}
S = \frac{1}{16 \pi G}\int \sqrt{-g} (R - 2\Lambda + P(R)\, F(\Box)\, Q(R)) d^4x ,
\end{equation}
where  $\Lambda$ is the cosmological constant, $P(R)$ and $Q(R)$ are some differentiable functions of $R$.
For a better insight into nonlocal effects, preliminary consideration of these models is usually without matter.
The most attention has been paid to the case when
$P(R) = Q(R) = R$, e.g. see \cite{biswas1,biswas2,koshelev,koshelev2,dimitrijevic1,dimitrijevic2,dimitrijevic3,dimitrijevic4,dimitrijevic5,dimitrijevic6,
dimitrijevic7,dimitrijevic8}.

Nonlocal gravity model which we investigate in this paper has $P(R) = Q(R) = \sqrt{R - 2 \Lambda}$ and action is given explicitly below in \eqref{2.1}.

\section{New Nonlocal Gravity Model}

\subsection{Action and Equations of Motion}

Our nonlocal model is given by the action
\begin{align}
  S =  \frac{1}{16\pi G} \int  \sqrt{R-2\Lambda} \, F(\Box) \, \sqrt{R-2\Lambda} \, \sqrt{-g}\; d^4x , \label{2.1}
 \end{align}
where  $F(\Box)= 1+ \mathcal{F}(\Box)= 1 + \displaystyle \sum_{n =1}^{\infty} f_{n}\, \Box^{n}$
and $\Box = \nabla_{\mu}\nabla^{\mu}= \frac{1}{\sqrt{-g}}\, \partial_\mu \, (\sqrt{-g}\, g^{\mu\nu}\, \partial_\nu )$ is the corresponding  d'Alembert operator.
  The action \eqref{2.1} can be rewritten in the form
\begin{align}
  S =  \frac{1}{16\pi G} \int ( R-2\Lambda + \sqrt{R-2\Lambda}\, \mathcal{F}(\Box)\,  \sqrt{R-2\Lambda}\, )\, \sqrt{-g}\; d^4x , \label{2.2}
 \end{align}
where there is separation on Einstein  $R - 2\Lambda$ part and nonlocal term \\ $\sqrt{R-2\Lambda}\, \mathcal{F}(\Box)\,  \sqrt{R-2\Lambda} .$

By variation of the action \eqref{2.1} with respect to the metric $g^{\mu\nu}$ we obtain the equations of motion for $g_{\mu\nu}$ tensor, i.e.
\begin{align}
&\bar{G}_{\mu\nu}= G_{\mu\nu}+ \Lambda g_{\mu\nu}   +    (R_{\mu\nu} - \nabla_{\mu}\nabla_{\nu} + g_{\mu\nu}\Box) \, V^{-1} \mathcal{F}(\Box) \, V \nonumber \\
 &+ \frac{1}{2} \sum_{n=1}^{+\infty} f_{n}\sum_{\ell =0}^{n-1}\Big
(g_{\mu\nu} \big(g^{\alpha\beta}\partial_{\alpha}\Box^{\ell} \, V \,
\partial_{\beta} \Box^{n-1-\ell} \, V + \Box^{\ell}\, V \Box^{n-\ell}\, V\big)   \nonumber \\
 &- 2 \partial_{\mu} \Box^{\ell}\, V \partial_{\nu}\Box^{n-1-\ell}\, V \Big )
 - \frac{1}{2}g_{\mu\nu} \, V \, \mathcal{F}(\Box) \, V = 0,     \label{2.3}
\end{align}
where $\bar{G}_{\mu\nu}$ is nonlocal modified Einstein tensor and $ V =  \sqrt{R-2\Lambda}.$ For detailed derivation of equations of motions \eqref{2.3} we refer to \cite{dimitrijevic9}.

By straightforward calculation we have checked that
\begin{align}
\triangledown^\mu\bar{G}_{\mu\nu} = 0  \, .       \label{2.3a}
\end{align}
Eq. \eqref{2.3} can be rewritten in the form
\begin{equation}
\bar{G}_{\mu\nu} = G_{\mu\nu}+ \Lambda g_{\mu\nu} - 8 \pi G \bar{T}_{\mu\nu} = 0 \, , \label{2.3b}
\end{equation}
where $ \bar{T}_{\mu\nu}$ can be considered as a nonlocal gravity imitation of the energy-momentum tensor in Einstein gravity.

In the sequel we are interested in finding some exact cosmological solutions of \eqref{2.3}.  When the metric is homogeneous and isotropic, i.e.
Friedmann-Lema\^{\i}tre-Robertson-Walker (FLRW) metric
\begin{align}
ds^2 = - dt^2 + a^2(t)\left(\frac{dr^2}{1-k r^2} + r^2 d\theta^2 + r^2 \sin^2 \theta d\phi^2\right), \quad (c=1) , \, \, k= 0, \pm 1 , \label{2.4}
\end{align}
 then
eq. \eqref{2.3} is equivalent to the following  two equations (trace and 00-component, respectively):
 \begin{align}
&4\Lambda - R  -2 V \, \mathcal{F}(\Box) \, V +  (R
 + 3 \Box )\, V^{-1} \mathcal{F}(\Box) \, V  \nonumber \\
 &+ \sum_{n=1}^{+\infty}f_{n}
\sum_{\ell =0}^{n-1} ( \partial_{\alpha}\Box^{\ell}\, V \partial^{\alpha} \Box^{n-1-\ell}\, V + 2\Box^{\ell}\, V \Box^{n-\ell}\, V)
  = 0,  \label{2.5}
\end{align}
\begin{align}
 &G_{00} - \Lambda    +   (R_{00} - \partial_{0}\partial_{0} - \Box) \, V^{-1} \mathcal{F}(\Box) \, V \nonumber \\
 &- \frac{1}{2} \sum_{n=1}^{+\infty} f_{n} \sum_{\ell =0}^{n-1}\Big( \partial_{\alpha}\Box^{\ell}\, V
\partial^{\alpha} \Box^{n-1-\ell}\, V + \Box^{\ell}\, V \Box^{n-\ell}\, V \nonumber \\
 &+ 2 \partial_{0} \Box^{\ell}\, V \partial_{0}\Box^{n-1-\ell}\, V \Big ) + \frac{1}{2} V \, \mathcal{F}(\Box) \, V = 0 ,  \label{2.6}
\end{align}
where $a(t)$ is the cosmic scale factor and
\begin{equation}
   R_{00} = - 3 \frac{\ddot{a}}{a} \, , \qquad   G_{00} = 3 \frac{\dot{a}^2  + k}{a^2} \,  .     \label{2.6a}
\end{equation}

The related Friedmann equations to \eqref{2.3b} are
\begin{equation}
\frac{\ddot{a}}{a} = - \frac{4\pi G}{3} (\bar{\rho} + 3 \bar{p}) + \frac{\Lambda}{3} \,,  \quad \frac{\dot{a}^2  + k}{a^2} = \frac{8\pi G}{3} \bar{\rho}
+ \frac{\Lambda}{3} \,,   \label{2.7}
\end{equation}
where $\bar{\rho}$ and $\bar{p}$ are analogs of the energy density and pressure of the dark side of the universe, respectively.
 Denote the corresponding equation of state as
$\bar{p} (t) = \bar{w}(t) \, \bar{\rho} (t) . $

\section{Cosmological Solutions}

For the FLRW metric \eqref{2.4} the corresponding scalar curvature is
\begin{align}
R (t) = 6\Big(\frac{\ddot a}{a} + \big(\frac{\dot a}{a}\big)^2 +\frac{k}{a^2}\Big).   \label{3.1}
\end{align}
 Operator $\Box$  acts on $R(t)$ as $\Box R = - \frac{\partial^2}{\partial t^2} {R}-3 H \frac{\partial}{\partial t} {R} ,$ where $H= \frac{1}{a} \frac{d a}{d t}\,  \equiv \frac{\dot{a}}{a}$ is the Hubble parameter.

Note that eqs. \eqref{2.5} and \eqref{2.6} do
 not allow the Minkowski space  solution, because $\Lambda = R = 0$ implies $V^{-1} = \infty .$
In the sequel we present some exact cosmological solutions with $\Lambda \neq 0 .$

  We shall use some cosmological parameters from Planck 2018 results \cite{Planck2018}  to test  validity of obtained solutions for the current state of the universe. The current Planck results for the $\Lambda$CDM universe are:
  \begin{itemize}
  \item  $H_0 = (67.40 \pm 0.50)$ km/s/Mpc  -- Hubble parameter;
  \item  $\Omega_m = 0.315 \pm 0.007$  -- matter density parameter;
  \item $\Omega_\Lambda = 0. 685$ -- $\Lambda$ density parameter;
  \item $t_0  = (13.801 \pm 0.024) \cdot 10 ^9$ yr -- age of the universe;
  \item $w_0 = - 1.03 \pm 0.03$  --  ratio of pressure to energy density.
  \end{itemize}

\subsection{Cosmological solution $ a (t) = A \, t^{\frac{2}{3}} \, e^{\frac{\Lambda}{14} t^2}  \,, \,\, k=0 $}

Scalar curvature \eqref{3.1} is
\begin{equation} \label{3.1.1}
R(t) = \frac{4}{3} t^{-2} + \frac{22}{7} \Lambda + \frac{12}{49}  \Lambda^2 t^2
\end{equation}
and  the Hubble parameter
\begin{align}
 H(t) = \frac{2}{3}  t^{-1} + \frac{1}{7} \Lambda t . \label{3.1.1a}
 \end{align}
   There is equality
\begin{align}
\Box \sqrt{R - 2 \Lambda} =  - \frac{3}{7} \Lambda  \sqrt{R - 2 \Lambda}    \label{3.1.2}
\end{align}
which leads to
\begin{align}
\mathcal{F}(\Box) \, \sqrt{R - 2 \Lambda} =  \mathcal{F}\big(- \frac{3}{7} \Lambda\big) \, \sqrt{R - 2 \Lambda} .  \label{3.1.3}
\end{align}
$R_{00}$ and $G_{00}$ are:
\begin{align}
R_{00} = \frac{2}{3} \, t^{-2} - \Lambda - \frac{3}{49}\Lambda^2 t^2 \,, \quad G_{00} =  \frac{4}{3} \, t^{-2} + \frac{4}{7} \Lambda + \frac{3}{49}\Lambda^2 t^2 \,.  \label{3.1.4}
\end{align}
Eqs. \eqref{2.5} and \eqref{2.6} are satisfied under conditions
\begin{align}
    \mathcal{F} \big(- \frac{3}{7}\Lambda \big) = - 1 \,,  \quad   \mathcal{F}' \big(- \frac{3}{7}\Lambda \big) = 0 \, , \quad \Lambda \neq 0 .  \label{3.1.5}
\end{align}

From \eqref{2.7} follows
\begin{equation}
\bar{\rho} (t) = \frac{2 t^{-2} + \frac{9}{98}\Lambda^2 t^2 - \frac{9}{14} \Lambda}{12 \pi G} \,,    \quad \bar{p}(t) = - \frac{\Lambda}{56 \pi G}
\big( \frac{3}{7} \Lambda t^2  - 1 \big)  .   \label{3.1.5a}
\end{equation}

 This cosmological solution $ a (t) = A \, t^{\frac{2}{3}} \, e^{\frac{\Lambda}{14} t^2} $ can be viewed as a product of
 $t^{\frac{2}{3}}$ factor, related to the matter dominated case in Einstein's gravity, and $e^{\frac{\Lambda}{14} t^2}$ which is related to an acceleration.
 Moreover, according to expression \eqref{3.1.1a},
 the Hubble parameter consists  of two terms, where  $\frac{2}{3}  t^{-1}$ is just $H(t)$ in Einstein's theory of gravity for the universe dominated by matter. The second term
 $\frac{1}{7} \Lambda t$ corresponds to an acceleration for $\Lambda > 0 .$ From \eqref{3.1.1a} follows that  term $\frac{2}{3}  t^{-1}$ is dominant for
 small $t$, while   $\frac{1}{7} \Lambda t$  plays leading role for larger times. Time dependent expansion acceleration is
 \begin{equation}
 \ddot{a}(t) = \big(- \frac{2}{9} t^{-2} + \frac{\Lambda}{3} + \frac{\Lambda^2}{49} t^{2}\big)\, a(t) . \label{3.1.6}
 \end{equation}
  Also, according to expressions \eqref{3.1.5a} follows that $\bar{w}(t)  \to - 1$ when $t \to \infty ,$ what corresponds to an analog of $\Lambda$ dark energy dominance    in the standard cosmological model.
 Therefore, one can say that nonlocal gravity model $\eqref{2.1}$ with cosmological solution
 $ a(t) = A\, t^{\frac{2}{3}} \, e^{\frac{\Lambda}{14} t^2} $
  describes some effects usually attributed to the dark matter and  dark energy. This solution is invariant under transformation $t \to - t$ and singular at cosmic time $t = 0$. Namely, $R (t), \, H(t)$ and   $\bar{\rho}(t)$ tend to $+\infty$  when $t \to 0,$ while $\bar{p}(0)$ is finite.

 Taking the above Planck results for  $t_0$  and $H_0$ in \eqref{3.1.1a}  one obtains $\Lambda = 1,05\cdot 10^{-35} \, s^{-2}$ (in $c=1$ units).  This is close to  $\Lambda = 0.98 \cdot 10^{-35} s^{-2}$ calculated by standard formula $\Lambda = 3 H_0^2 \Omega_\Lambda$. From \eqref{3.1.1a} one can also calculate time ($t_m$) for which the Hubble parameter has minimum value $H_m$, i.e. $t_m = 21,1 \cdot 10^9$ yr and $H_m = 61, 72$ km/s/Mpc.

 According to \eqref{3.1.6}, beginning of the universe expansion acceleration was at $t_a = 7,84 \cdot 10^{9}$ yr, or in other words at $5.96$ billion years ago.

 From the Friedmann  equation $\frac{\dot{a}^2}{a^2} = \frac{8\pi G}{3} \bar{\rho}
+ \frac{\Lambda}{3}$, combined with expression \eqref{3.1.1a} for the Hubble parameter,   one can calculate the critical energy density $\rho_c$ and the energy density of the dark matter $\bar{\rho}$  for the solution $ a(t) = A\, t^{\frac{2}{3}} \, e^{\frac{\Lambda}{14} t^2} $:
\begin{align}
\label{3.1.7}
\rho_c &= \frac{3}{8 \pi G} H_0^2 = 8,51 \cdot 10^{-30} \frac{g}{cm^3} \\
\bar{\rho} &= \Big(\frac{4}{9} t_0^{-2}  - \frac{\Lambda}{7} + \frac{\Lambda^2}{49} t_0^2 \Big) \frac{3}{8 \pi G} = 2,26 \cdot 10^{-30} \, \frac{g}{cm^3}. \label{3.1.8}
\end{align}
It follows that $\bar{\Omega} = \frac{\bar{\rho}}{\rho_c}  = 0,265$. Since $\Omega_v$   for the visible matter is approximatively  $\Omega_v = 0,05 $, then  $\bar\Omega_\Lambda = 1 - \bar{\Omega} - \Omega_v = 0, 685 .$

\subsection{Cosmological solution $ a (t) = A  \, e^{\frac{\Lambda}{6} t^2} \,, \,\, k=0  $}

In this case scalar curvature \eqref{3.1} is
\begin{equation} \label{3.2.1}
R(t) = 2 \Lambda (1 + \frac{2}{3} \Lambda t^2)
\end{equation}
 and $\,
 H(t) = \frac{1}{3} \Lambda t \, ,  \quad \Box \sqrt{R - 2 \Lambda} = - \frac{2}{\sqrt{3}} \Lambda |\Lambda| |t|$. It follows a useful equality
\begin{align}
\Box \sqrt{R - 2 \Lambda} =  -\Lambda  \sqrt{R - 2 \Lambda}    \label{3.2.2}
\end{align}
which significantly simplifies  analysis of equations of motion \eqref{2.5} and \eqref{2.6}. From \eqref{3.2.2} follows
\begin{align}
&\Box^n \, \sqrt{R - 2 \Lambda}  = (- \Lambda)^n \, \sqrt{R - 2 \Lambda} \,, \quad n \geq 0 ,  \label{3.2.3} \\
&\mathcal{F}(\Box) \, \sqrt{R - 2 \Lambda} =  \mathcal{F}(- \Lambda) \, \sqrt{R - 2 \Lambda} .  \label{3.2.4}
\end{align}
Calculation of $R_{00}$ and $G_{00}$ gives
\begin{align}
R_{00} = -\frac{\Lambda^2}{3} \, t^2 - \Lambda \,, \quad G_{00} =  \frac{\Lambda^2}{3} \, t^2  \,.  \label{3.2.5}
\end{align}
Equations of motion \eqref{2.5} and \eqref{2.6} are satisfied by this solution if and only if
\begin{align}
    \mathcal{F} (- \Lambda) = \sum_{n=1}^{+\infty} f_n (-\Lambda)^n = - 1 \,,  \quad   \mathcal{F}' (- \Lambda) =
    \sum_{n=1}^{+\infty} f_n \, n (-\Lambda)^{n-1} = 0.  \label{3.2.6}
\end{align}

According to \eqref{2.7} follows
\begin{equation}
\bar{\rho}(t) = \frac{\Lambda}{8\pi G} \big(\frac{\Lambda}{3} t^2 -1  \big) \,, \quad \bar{p}(t) = - \frac{\Lambda}{24 \pi G} \big(\Lambda t^2  - 1 \big) . \label{3.2.7}
\end{equation}

Solution  $a (t) = A  \, e^{\frac{\Lambda}{6} t^2}$ is nonsingular with $R(0) = 2 \Lambda$ and $H (0) = 0 .$ There is acceleration expansion
 $\ddot{a}(t) = \big(\frac{\Lambda}{3} + \frac{\Lambda^2}{9} \, t^2 \big) \, a(t)$ which is positive and  increasing  with time.

\subsection{Cosmological solution $ a (t) = A  \,  e^{\pm \sqrt{\frac{\Lambda}{6}}\, t} \,, \, \, \Lambda > 0 \,, \,\, k=\pm 1 $}

Now
\begin{equation} \label{3.3.1}
R(t) = \frac{6 k}{A^2} e^{\mp  \sqrt{\frac{2}{3}\Lambda} \, t}  + 2 \Lambda ,
\end{equation}
$H = \pm \sqrt{\frac{\Lambda}{6}}$ and useful equality is $\Box \sqrt{R - 2 \Lambda} = \frac{\Lambda}{3} \sqrt{R - 2 \Lambda} .$
Also we have
\begin{equation}
R_{00} = - \frac{\Lambda}{2} ,        \qquad   G_{00} =  \frac{3 k}{A^2} e^{\mp  \sqrt{\frac{2}{3}\Lambda} \, t}  + \frac{\Lambda}{2} . \label{3.3.2}
\end{equation}
Equations of motion \eqref{2.5} and \eqref{2.6} are satisfied if and only if
\begin{equation} \label{3.3.3}
\mathcal{F} \big(\frac{\Lambda}{3}\big) = -1 \, , \qquad   \mathcal{F}' \big(\frac{\Lambda}{3}\big) = 0 .
\end{equation}

Related $\bar{\rho}$ and $\bar{p}$  are
\begin{equation}
\bar{\rho}(t) = \frac{ -\frac{\Lambda}{2} + \frac{3 k}{A^2} \,  e^{\mp \sqrt{\frac{2}{3} \Lambda} \, t}}{8 \pi G} \,,    \quad
\bar{p}(t)  = \frac{ \frac{\Lambda}{2} - \frac{ k}{A^2} \,  e^{\mp \sqrt{\frac{2}{3} \Lambda} \, t}}{8 \pi G}  . \label{3.3.4}
\end{equation}

In this case, we have two solutions:  $ (i)\, \, a (t) = A  \,  e^{\sqrt{\frac{\Lambda}{6}}\, t}$ and $ (ii) \, \,
 a (t) = A  \,  e^{-\sqrt{\frac{\Lambda}{6}}\, t}$, for both $k = +1$ and $k = - 1$. They are similar to the de Sitter solution $ a (t) = A  \,  e^{\pm \sqrt{\frac{\Lambda}{3}}\, t} \,,  \, k= 0 ,$ but have time dependent $R(t), \bar{\rho} (t)$ and $\bar{p} (t) .$  When $t \to +\infty ,$
 parameter $\bar{w}(t) \to - 1$ in the case $(i)$ and  $\bar{w}(t) \to - \frac{1}{3}$  for solution  $(ii)$.

\bigskip

The above solutions have time dependent scalar curvature $R(t) .$  Below we  present some cosmological solutions with constant $R(t) = R_0 .$

\subsection{Cosmological solutions with $R_0 = 4 \Lambda >  0$}

There are three cases as in Einstein's gravity:

{\it (i)\, Case} $a (t) = A  \,  e^{\pm \sqrt{\frac{\Lambda}{3}}\, t} \,, \,\, k=0 .$ $H(t) = \pm \sqrt{\frac{\Lambda}{3}} $.

{\it (ii)\, Case} $ a (t) = \sqrt{\frac{3}{\Lambda}} \cosh{\sqrt{\frac{\Lambda}{3}}\, t} \,, \,\, k=+1 .$ Now $H(t) = \sqrt{\frac{\Lambda}{3}} \tanh{\sqrt{\frac{\Lambda}{3}}\, t} , \, \, G_{00} = \Lambda .$

{\it (iii)\, Case} $ a (t) = \sqrt{\frac{3}{\Lambda}} \left|\sinh{\sqrt{\frac{\Lambda}{3}}\, t}\right| \,, \,\, k=-1 .$ Here $H(t) = \sqrt{\frac{\Lambda}{3}} \coth{\sqrt{\frac{\Lambda}{3}}\, t} , \, \, G_{00} = \Lambda .$

  Equations of motion \eqref{2.5} and \eqref{2.6} are satisfied without conditions on function $\mathcal{F}(\Box) ,$ because
$\Box \sqrt{R - 2 \Lambda} = 0 .$

\subsection{Cosmological solution with $R_0 = -4 |\Lambda| <  0$}

The corresponding solution has the form $a (t) = \sqrt{-\frac{3}{\Lambda}} \left| \cos{\sqrt{-\frac{\Lambda}{3}}\, t}\right|  ,$ where $\Lambda$ is negative cosmological constant. In this case $H(t)= - \sqrt{-\frac{\Lambda}{3}} \tan{\sqrt{-\frac{\Lambda}{3}}\, t} ,$   $\, G_{00} =- |\Lambda|$ and $k = -1 .$

\section{Concluding Remarks}

In this paper, we presented a few exact cosmological solutions for nonlocal gravity model given by  \eqref{2.1}.
These solutions are valid for  $\Lambda \neq 0 $ and without matter. Some of the solutions are not contained in Einstein's gravity with  cosmological constant $\Lambda$. In particular, solution $ a (t) = A \, t^{\frac{2}{3}} \, e^{\frac{\Lambda}{14} t^2}$ deserves further investigation, because it imitates some effects which are usually attributed to the dark matter and the dark energy. Calculated cosmological parameters are in good agreement with observations as well.
We plan to  investigate also other phenomenological aspects according to physical foundations of cosmology \cite{muhanov}.

Cosmological relevance of the other above presented solutions will be considered elsewhere.

In nonlocal gravity model  \eqref{2.1}, analytic function $\mathcal{F}(\Box)$ is rather arbitrary -- it is so far
constrained only by a few conditions. Using procedure presented in our paper \cite{dimitrijevic4}, one can show
that there exists analytic function $\mathcal{F}(\Box)$ with  the de Sitter background without a ghost and
tachyon, and will be presented elsewhere.

The following important remark follows from the analysis of physical excitations around the
de Sitter background performed in \cite{dimitrijevic4}, subsection 5.1.
Our model (3) modifies only the scalar sector of perturbations, i.e.
spin-0 perturbations, e.g. Newtonian potential may get modified. Tensor, i.e.
 spin-2 modes remain the same as in general relativity with overall global constant
factor renormalization at most. This follows from  equations (57) and (58) for the
second order variation of the action presented in \cite{dimitrijevic4}.
\begin{figure}[h]
			\begin{center}
				\includegraphics[width=9 cm]{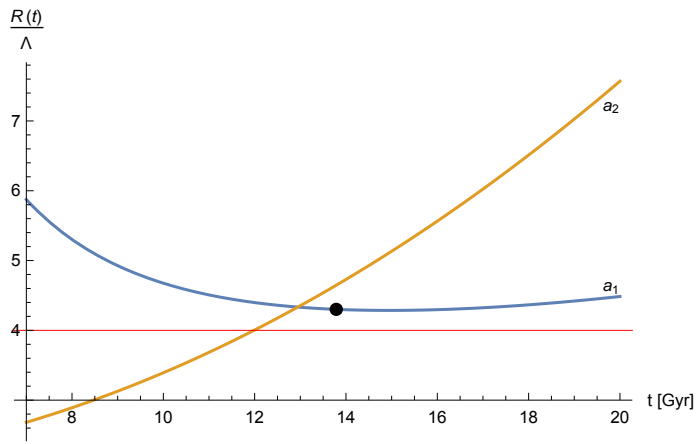}
				\caption{Graphics of the Ricci scalar $R(t)/\Lambda$ for three cases:  a$_1$ line for case \eqref{3.1.1}, a$_2$ line for case \eqref{3.2.1}  and horizontal line for the de Sitter case. Line a$_1$ is almost constant in a wide region around present cosmic time $t_0 = 13.8 \cdot 10^9$ yr. It has value $4.30$ at $t_0$ (denoted by a dot), what is  $7.5 \%$ larger than $4$ (de Sitter case). Its minimum value is $4.28$ at $t_1 = 14.9 \cdot 10^9$ yr, i.e. $7 \%$ above $4$.}
				\label{curvature2}
			\end{center}	
			\end{figure}
For our most
important solution, i.e. $ a (t) = A \, t^{\frac{2}{3}} \, e^{\frac{\Lambda}{14} t^2}$, the Ricci
scalar \eqref{3.1.1} is almost constant in a wide region around the present cosmic time and about $7.5\%$ is larger
than $\Lambda = 1.05 \cdot 10^{-35} s^{-2}$ (see Fig. \ref{curvature2}).
Consequently,  the model considered in this paper contains two solutions which  practically do  not modify what is
related to gravitons and in particular do not change the
speed of gravitational waves comparing to general relativity with a
 cosmological constant. The speed of light in the de Sitter space is considered in \cite{pereira}.

At the end, note that cosmological solutions with nonlocal dynamics of a scalar field inspired by string field theory and $p$-adic string theory have been considered in the matter sector of Einstein equations, e.g. see \cite{arefeva1,arefeva2} and references therein.

\section*{Acknowledgements}
Work on this paper was partially supported by the Ministry of Education, Science and Technological Development of  Republic of Serbia, grant No 174012. The authors thank reviewer of this paper for valuable comments.

\end{document}